\input{aipcheck}

\documentclass[
    ,final            
  ]
  {aipproc}

\layoutstyle{6x9}

\usepackage{psfrag}
\newcommand{\sss}{\scriptscriptstyle}

\begin{document}

\title{NLO corrections to the $\Delta F=2$ Hamiltonian in the MSSM with non-degenerate squarks}


\author{Javier Virto}{
  address={INFN, Sezione di Roma, I-00185 Rome, Italy} }

\begin{abstract}
We present the next-to-leading strong interaction corrections to the $\Delta F=2$ Hamiltonian in the MSSM with exact diagonalization of the squark mass matrices. These results allow phenomenological studies of neutral meson mixing in scenarios with non-degenerate squarks, with control over the renormalization scale and scheme dependence.
\end{abstract}

\maketitle


Flavor changing low-energy processes such as weak meson decays or the mixing of neutral mesons, are unique for indirect new physics searches because they are very sensitive to heavy degrees of freedom in loop amplitudes. This is due to the suppression of flavor-changing neutral currents in the SM, in contrast to the enormous flavor violation typical of new physics scenarios with generic flavor structure. Precise measurements --specially those related to $K$ and $B$ mesons-- have put stringent constraints on either the new physics scale or the flavor non-universality of the new couplings \cite{Antonelli:2009ws}. This is explicitly clear in SUSY models, and in particular in the MSSM, which is the paradigmatic extension of the SM.

In the MSSM, the leading source of flavor violation arises from a misalignment between quark and squark mass eigenbases induced by soft SUSY-breaking terms: in the basis for the superfields in which the quark mass matrices are diagonal (the so called super-CKM basis), the squark mass matrices contain non-zero non-diagonal elements (mass insertions), which are responsible for the flavor transition. Extensive phenomenological studies have been performed which constrain strongly the size of these mass insertions, based on FCNC processes \cite{Ciuchini:1998ix,Becirevic:2001jj,Silvestrini:2007yf,Ciuchini:2007cw,Crivellin:2009ar}, vacuum stability requirements \cite{hep-ph/9507294,hep-ph/9606237} and charged-current processes \cite{arXiv:0810.1613}. 

However, the increasing precision with which these quantities must be determined requires theoretical calculations of flavor violating processes at increasing order in perturbation theory, as well as more precise determinations of non-perturbative parameters. Here we focus on the calculation of matching conditions (Wilson coefficients at the SUSY scale) for the $\Delta F=2$ Hamiltonian, which are required for the theoretical determination of $K-\bar K$, $D-\bar D$ and $B_{d,s}-\bar B_{d,s}$ mixing amplitudes. The case of $B_s-\bar B_s$ mixing has recently caused some excitement in relation with a deviation of the measured mixing phase relative to the SM expectation \cite{arXiv:0712.2397,arXiv:0802.2255,arXiv:0803.0659}, and if such deviation is confirmed, a precise calculation of that phase in different models will be very convenient.

\begin{figure}[t]
 \centering
 \psfrag{sb}{\small $\bar s$}\psfrag{bb}{\small $\bar b$}
 \psfrag{s}{\small $s$}\psfrag{b}{\small $b$} \psfrag{a}{\small $(a)$} \psfrag{c}{\small $(b)$}
 \psfrag{g}{\small $\tilde g$}\psfrag{qi}{\small $\tilde q_i$} \psfrag{qj}{\small $\tilde q_j$}
  \includegraphics[height=3.5cm,width=4.5cm]{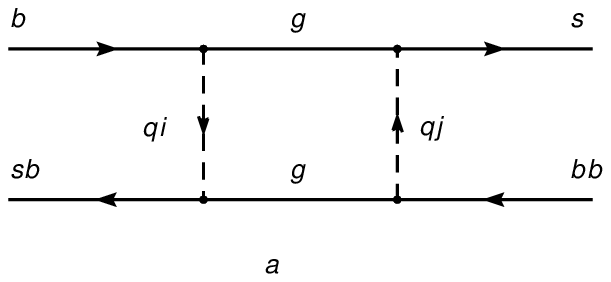}\hspace{1cm}
  \includegraphics[height=3.5cm,width=9cm]{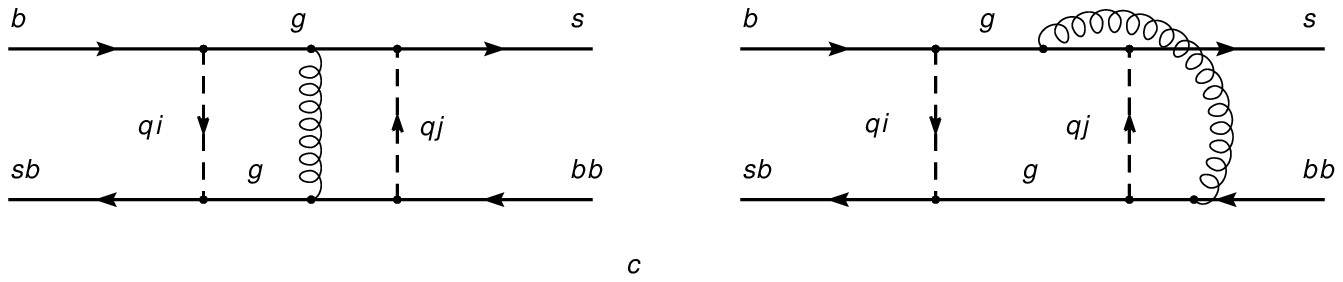}
 \caption{A sample of the diagrams contributing to the MSSM matching conditions for $B_s-\bar B_s$ mixing at (a) leading order, and (b) next-to-leading order in $\alpha_s$.}
\label{fig1}
\end{figure}

Leading order (LO) strong interaction matching conditions in the MSSM have been computed in Refs.~\cite{Gerard:1984bg,hep-ph/9604387,Hagelin:1992tc}, and arise from squark-gluino box diagrams (Fig.~\ref{fig1}.a). The corresponding next-to-leading order (NLO) corrections arise from two loop diagrams such as those shown in Fig.~\ref{fig1}.b, and have been computed in Ref.~\cite{hep-ph/0606197} within the degenerate Mass Insertion Approximation (that is, at the leading order in an expansion in mass insertions, and assuming that the diagonal elements of the squark mass matrix are equal). Here we report on the calculation of the NLO strong interaction corrections for non-degenerate squark masses, that is, with exact diagonalization of the squark mass matrices. This calculation was performed in Ref.~\cite{Virto:2009wm}, extending the known results beyond the Mass Insertion Approximation (MIA), and the full expressions for the Wilson coefficients can be found in that paper.

These NLO corrections valid for non-degenerate squark masses are important for two reasons:

\begin{enumerate}

\item At LO, the renormalization scale and scheme cannot be specified for the strong coupling, leading to a large uncertainty related to scale and scheme ambiguities. At NLO, however, scheme independent results and NLO scale invariance are accomplished. The reduction of this uncertainty is numerically important due to the large anomalous dimensions of the $\Delta F=2$ operators involved. The analysis of Ref.~\cite{hep-ph/0606197} shows that this uncertainty is reduced from 10-15\% down to a few percent.  

\item Scenarios with large squark mass splittings can give rise to a quite different pattern of correlations between $\Delta F=1$ and $\Delta F=2$ observables, such that $\Delta F=2$ processes can be affected very differently by the experimental bounds on $\Delta F=1$ processes. An example of how this comes about in a hierarchical scenario is given in Ref.~\cite{arXiv:0812.3610}, where it is shown that the bounds from $B\to X_s\gamma$ can be partially evaded allowing a large $B_s$ mixing phase.

\end{enumerate}

From the results with exact diagonalization of the squark mass matrices presented here and in Ref.~\cite{Virto:2009wm}, it is possible to recover the Wilson coefficients in the MIA. This is done by relating the rotation matrices to the mass insertions, expanding to lowest order in the mass insertion expansion and taking the limit of equal diagonal elements of the mass matrices. The corresponding expressions agree with the results of Ref.~\cite{hep-ph/0606197}, verifying the correctness of their calculation. Moreover, in order to provide a clear comparison between the degenerate and non-degenerate scenarios, it is convenient to expand the exact results to lowest order in the mass insertion expansion but keeping the squark masses non-degenerate. This scheme might be called the \emph{non-degenerate mass insertion approximation} (NDMIA). 

In order to get an idea of the size of the new contributions, we consider a particular scenario in the NDMIA for the case of $B_s-\bar B_s$ mixing. This scenario is based on the ``hierarchical'' setup of Ref.~\cite{arXiv:0812.3610}, in which the first two generation squarks are given a common mass $\tilde m_{12}$, different from a common mass $\tilde m_3$ for the third generation squarks. Moreover, $\tilde m_3$ is assumed to be near the electroweak scale in order to ensure naturalness, while $\tilde m_{12}$ could be well above (thus reducing the conflict with kaon physics data, for example).

\begin{figure}
\centering
\psfrag{xh}{$x_h$} \psfrag{C2NLO}{\hspace{-0.5cm}$\stackrel{1\ +\ C_2^{\rm \sss NLO}/C_2^{\rm \sss LO}}{}$}
\psfrag{C3NLO}{$\stackrel{\hspace{-0.5cm}1\ +\ C_3^{\rm \sss NLO}/C_3^{\rm \sss LO}}{}$}
\includegraphics[width=6cm]{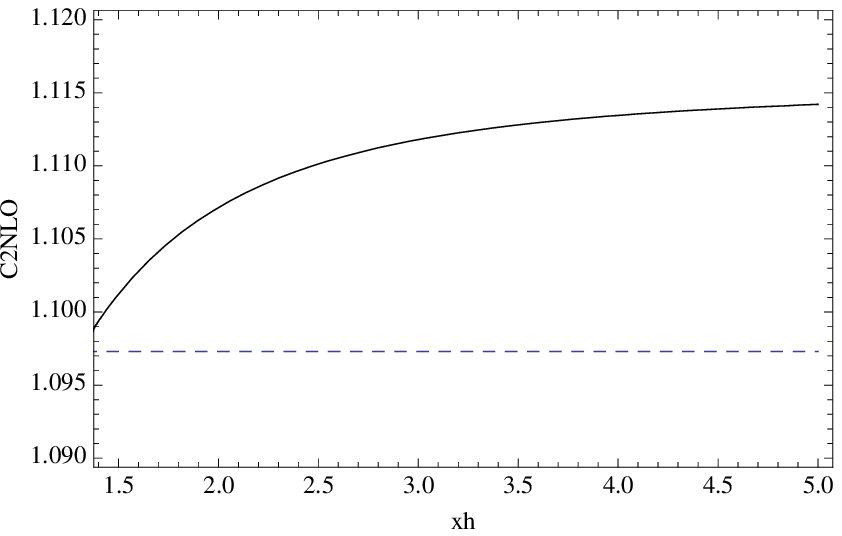}\qquad\quad
\includegraphics[width=6cm]{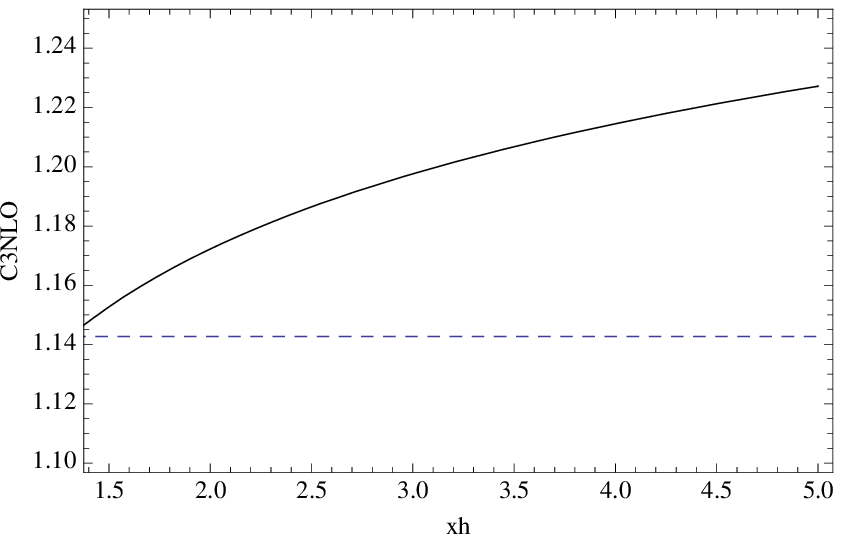}
\vspace{-0.5cm}
\caption{Relative importance of the full NLO result with respect to the LO, for the Wilson coefficients $C_2$ and $C_3$, as a function of $x_h\equiv \tilde m_{12}^2/m_{\tilde g}^2$, where $\tilde m_{12}$ is a common mass for first and second generation squarks. The dashed lines correspond to the degenerate (MIA) scenario. We have chosen $400~{\rm GeV}$ for third generation squark masses and $m_{\tilde g}=\mu=350~{\rm GeV}$, where $\mu$ is the matching scale.}
\label{fig2}
\end{figure}

The plots in Fig.~\ref{fig2} illustrate the relative importance of the NLO corrections, as a function of the mass splitting between the light and heavy squarks. The NLO correction is typically a $\sim 10\%$ effect in the degenerate case, but its importance increases with the mass splitting. For heavy squarks of about a TeV, the NLO contribution to $C_3$ can be up to a $\sim 25\%$ correction.

\begin{figure}
\centering
\psfrag{xl}{$x_l$} \psfrag{C1NLO}{\hspace{-0.5cm}\small $C_1^{\rm \sss NLO}\ {\scriptstyle (\times 10^{8})}$}
\psfrag{C4NLO}{\hspace{-0.5cm}\small $C_4^{\rm \sss NLO}\ {\scriptstyle (\times 10^{6})}$}
\psfrag{C5NLO}{\hspace{-0.5cm}\small $C_5^{\rm \sss NLO}\ {\scriptstyle (\times 10^{7})}$}
\includegraphics[width=4.5cm,height=3.5cm]{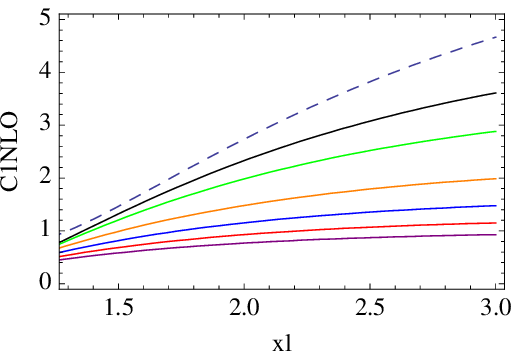}\quad
\includegraphics[width=4.5cm,height=3.5cm]{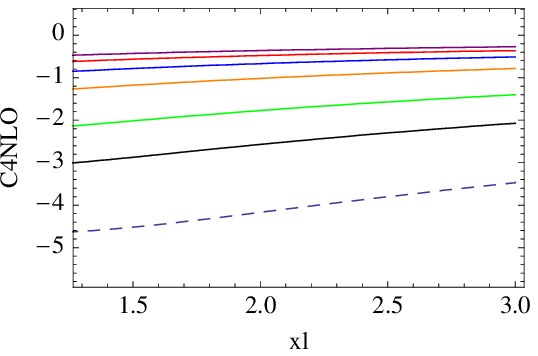}\quad
\includegraphics[width=4.5cm,height=3.5cm]{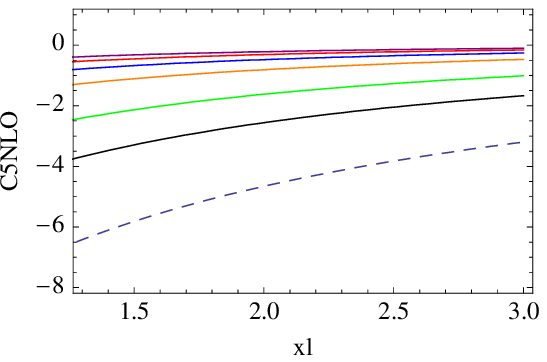}
\vspace{-0.5cm}
\caption{NLO Wilson coefficients $C_1^{NLO}$, $C_4^{NLO}$ and $C_5^{NLO}$ as a function of $x_l\equiv \tilde m_{3}^2/m_{\tilde g}^2$, where $\tilde m_{3}$ is a common mass for third generation squarks. The plots are in units of $(\alpha_s^3/\pi) \delta_{LL}^2$ and $(\alpha_s^3/\pi) \delta_{LL}\delta_{RR}$ for $C_1$ and $C_{4,5}$ respectively. The different lines correspond to different values of $x_h\equiv \tilde m_{12}^2/m_{\tilde g}^2$, and range from the MIA case, $x_h=x_l$ (dashed) to $x_h=1.5 x_l, 2 x_l,3 x_l, 4 x_l, 5 x_l, 6 x_l$.}
\label{fig3}
\end{figure}

In order to analyze more closely the role of the mass splittings in the NLO corrections, we consider $C_1, C_4$ and $C_5$ as a function $x_l\equiv\tilde m_3^2/m_{\tilde g}^2$, for different splittings between $x_l$ and $x_h=\tilde m_{12}^2/m_{\tilde g}^2$. This is shown in Fig.~\ref{fig3}, where the dashed lines correspond to $x_l=x_h$ (that is, the degenerate case), and --departing smoothly from that limit-- the solid lines show increasing values of $x_h/x_l$. In these plots we take the dimensionless mass insertions $\delta_{L,R}=\delta_{R,L}=0$ and $m_{\tilde g}=\mu=350~{\rm GeV}$. We see that increasing the heavy scale tends to reduce systematically the size of the NLO contribution, being largest in the degenerate case.

However, caution must be taken when interpreting these results, since the NLO Wilson coefficients are renormalization scale and scheme dependent. These results correspond to the renormalization scheme adopted in Ref.~\cite{Virto:2009wm} (NDR with some subtleties), and will vary in other schemes. Of course, the scheme dependence cancels in physical amplitudes (against the scheme dependence of the matrix elements of the operators), so the true impact of these NLO corrections can only be established by analyzing their effect on observables. A full phenomenological analysis of these corrections and their impact on the bounds on the mass insertions (beyond the mass insertion approximation) will be presented in the future.

\begin{theacknowledgments}
I would like to thank Luca Silvestrini, Enrico Franco and Marco Ciuchini for discussions on topics related to this work. I would like to thank as well the organizers and the attendees to the SUSY'09 conference for a very stimulating workshop. ~J.V. is associated to the Dipartimento di Fisica, Universita di Roma `La Sapienza'. 
\end{theacknowledgments}





\bibliographystyle{aipproc}   


\IfFileExists{\jobname.bbl}{}
 {\typeout{}
  \typeout{******************************************}
  \typeout{** Please run "bibtex \jobname" to optain}
  \typeout{** the bibliography and then re-run LaTeX}
  \typeout{** twice to fix the references!}
  \typeout{******************************************}
  \typeout{}
 }



\end{document}